\documentclass[twocolumn,showpacs,preprintnumbers,superscriptaddress,amsmath,amssymb,prl,aps,floatfix]{revtex4}

\usepackage{graphicx}
\usepackage{dcolumn}
\usepackage{bm}

\begin{document}  
\newcommand{\Done}{$D_1 (2420)^0$}
\newcommand{\D}{$D_J^{(*)}$}
\newcommand{\functionOne}{$A(1+B* \cos^2 \theta$}
\newcommand{\functionTwo}{$A \sin^2 \theta$}
\newcommand{\Dtwo}{$D_2 (2460)^0$}
\newcommand{\Dstp}{$D^{*+} \pi^-_{decay}$}
\newcommand{\ra}{\rightarrow}
\newcommand{\dstdeckpi}{$D^{*+} \rightarrow D^0 \pi^+_{slow}, ~D^0 \rightarrow K^- \pi^+$}

\newcommand{\chicut}{16.}
\newcommand{\lxycut}{3.}
\newcommand{\resolution}{$6~GeV/c^2$}

\newcommand{\dnol}{$D^0$}
\newcommand{\dd}{$D_J^{(*)}$}
\newcommand{\ddchain}{$D_J^{(*)} \rightarrow D^{*+} \pi_{dec}^-, ~D^{*+} \rightarrow D^0 \pi^+_{slow}, ~D^0 \rightarrow \
K^- \pi^+$}
\newcommand{\ipb}{\ensuremath{\mathrm{pb^{-1}}}}
\newcommand{\ifb}{\ensuremath{\mathrm{fb^{-1}}}}
\newcommand{\TeV}{\ensuremath{\matqq    hrm{Te\kern -0.1em V}}}
\newcommand{\TeVc}{\ensuremath{\mathrm{Te\kern -0.1em V\!/}c}}
\newcommand{\TeVcc}{\ensuremath{\mathrm{Te\kern -0.1em V\!/}c^2}}
\newcommand{\GeV}{\ensuremath{\mathrm{Ge\kern -0.1em V}}}
\newcommand{\GeVc}{\ensuremath{\mathrm{Ge\kern -0.1em V\!/}c}}
\newcommand{\GeVcc}{\ensuremath{\mathrm{Ge\kern -0.1em V\!/}c^2}}
\newcommand{\MeV}{\ensuremath{\mathrm{Me\kern -0.1em V}}}
\newcommand{\MeVc}{\ensuremath{\mathrm{Me\kern -0.1em V\!/}c}}
\newcommand{\MeVcc}{\ensuremath{\mathrm{Me\kern -0.1em V\!/}c^2}}
\newcommand{\Tesla}{\ensuremath{\mathrm{T}}}
\newcommand{\cm}{\ensuremath{\mathrm{cm}}}
\newcommand{\um}{\ensuremath{\mathrm{\mu m}}}
\newcommand{\ps}{\ensuremath{\mathrm{ps}}}
\newcommand{\cdfii}{CDF\,II~}
\newcommand{\ppbar}{p\overline{p}}
\newcommand{\at}{\symbol{64}}
\newcommand{\sye}[1]{\ensuremath{~\pm #1}}
\newcommand{\ase}[2]{\ensuremath{^{~+ #1}_{~- #2}}}
\newcommand{\asi}[2]{\ensuremath{^{~- #1}_{~+ #2}}}

\newcommand{\myto}{\kern -0.3em\to\kern -0.2em}
\newcommand{\Lxy}{L_{\rm xy}}
\newcommand{\BR}{{\mathrm {BR}}}

\title{
Measurement of mass and width of the excited charmed meson states
$D_1^0$ and $D_2^{*0}$ at CDF}
\affiliation{Institute of Physics, Academia Sinica, Taipei, Taiwan 11529, Republic of China} 
\affiliation{Argonne National Laboratory, Argonne, Illinois 60439} 
\affiliation{Institut de Fisica d'Altes Energies, Universitat Autonoma de Barcelona, E-08193, Bellaterra (Barcelona), Spain} 
\affiliation{Baylor University, Waco, Texas  76798} 
\affiliation{Istituto Nazionale di Fisica Nucleare, University of Bologna, I-40127 Bologna, Italy} 
\affiliation{Brandeis University, Waltham, Massachusetts 02254} 
\affiliation{University of California, Davis, Davis, California  95616} 
\affiliation{University of California, Los Angeles, Los Angeles, California  90024} 
\affiliation{University of California, San Diego, La Jolla, California  92093} 
\affiliation{University of California, Santa Barbara, Santa Barbara, California 93106} 
\affiliation{Instituto de Fisica de Cantabria, CSIC-University of Cantabria, 39005 Santander, Spain} 
\affiliation{Carnegie Mellon University, Pittsburgh, PA  15213} 
\affiliation{Enrico Fermi Institute, University of Chicago, Chicago, Illinois 60637} 
\affiliation{Joint Institute for Nuclear Research, RU-141980 Dubna, Russia} 
\affiliation{Duke University, Durham, North Carolina  27708} 
\affiliation{Fermi National Accelerator Laboratory, Batavia, Illinois 60510} 
\affiliation{University of Florida, Gainesville, Florida  32611} 
\affiliation{Laboratori Nazionali di Frascati, Istituto Nazionale di Fisica Nucleare, I-00044 Frascati, Italy} 
\affiliation{University of Geneva, CH-1211 Geneva 4, Switzerland} 
\affiliation{Glasgow University, Glasgow G12 8QQ, United Kingdom} 
\affiliation{Harvard University, Cambridge, Massachusetts 02138} 
\affiliation{Division of High Energy Physics, Department of Physics, University of Helsinki and Helsinki Institute of Physics, FIN-00014, Helsinki, Finland} 
\affiliation{University of Illinois, Urbana, Illinois 61801} 
\affiliation{The Johns Hopkins University, Baltimore, Maryland 21218} 
\affiliation{Institut f\"{u}r Experimentelle Kernphysik, Universit\"{a}t Karlsruhe, 76128 Karlsruhe, Germany} 
\affiliation{High Energy Accelerator Research Organization (KEK), Tsukuba, Ibaraki 305, Japan} 
\affiliation{Center for High Energy Physics: Kyungpook National University, Taegu 702-701; Seoul National University, Seoul 151-742; and SungKyunKwan University, Suwon 440-746; Korea} 
\affiliation{Ernest Orlando Lawrence Berkeley National Laboratory, Berkeley, California 94720} 
\affiliation{University of Liverpool, Liverpool L69 7ZE, United Kingdom} 
\affiliation{University College London, London WC1E 6BT, United Kingdom} 
\affiliation{Massachusetts Institute of Technology, Cambridge, Massachusetts  02139} 
\affiliation{Institute of Particle Physics: McGill University, Montr\'{e}al, Canada H3A~2T8; and University of Toronto, Toronto, Canada M5S~1A7} 
\affiliation{University of Michigan, Ann Arbor, Michigan 48109} 
\affiliation{Michigan State University, East Lansing, Michigan  48824} 
\affiliation{Institution for Theoretical and Experimental Physics, ITEP, Moscow 117259, Russia} 
\affiliation{University of New Mexico, Albuquerque, New Mexico 87131} 
\affiliation{Northwestern University, Evanston, Illinois  60208} 
\affiliation{The Ohio State University, Columbus, Ohio  43210} 
\affiliation{Okayama University, Okayama 700-8530, Japan} 
\affiliation{Osaka City University, Osaka 588, Japan} 
\affiliation{University of Oxford, Oxford OX1 3RH, United Kingdom} 
\affiliation{University of Padova, Istituto Nazionale di Fisica Nucleare, Sezione di Padova-Trento, I-35131 Padova, Italy} 
\affiliation{University of Pennsylvania, Philadelphia, Pennsylvania 19104} 
\affiliation{Istituto Nazionale di Fisica Nucleare Pisa, Universities of Pisa, Siena and Scuola Normale Superiore, I-56127 Pisa, Italy} 
\affiliation{University of Pittsburgh, Pittsburgh, Pennsylvania 15260} 
\affiliation{Purdue University, West Lafayette, Indiana 47907} 
\affiliation{University of Rochester, Rochester, New York 14627} 
\affiliation{The Rockefeller University, New York, New York 10021} 
\affiliation{Istituto Nazionale di Fisica Nucleare, Sezione di Roma 1, University di Roma ``La Sapienza," I-00185 Roma, Italy} 
\affiliation{Rutgers University, Piscataway, New Jersey 08855} 
\affiliation{Texas A\&M University, College Station, Texas 77843} 
\affiliation{Istituto Nazionale di Fisica Nucleare, University of Trieste/\ Udine, Italy} 
\affiliation{University of Tsukuba, Tsukuba, Ibaraki 305, Japan} 
\affiliation{Tufts University, Medford, Massachusetts 02155} 
\affiliation{Waseda University, Tokyo 169, Japan} 
\affiliation{Wayne State University, Detroit, Michigan  48201} 
\affiliation{University of Wisconsin, Madison, Wisconsin 53706} 
\affiliation{Yale University, New Haven, Connecticut 06520} 
\author{A.~Abulencia}
\affiliation{University of Illinois, Urbana, Illinois 61801}
\author{D.~Acosta}
\affiliation{University of Florida, Gainesville, Florida  32611}
\author{J.~Adelman}
\affiliation{Enrico Fermi Institute, University of Chicago, Chicago, Illinois 60637}
\author{T.~Affolder}
\affiliation{University of California, Santa Barbara, Santa Barbara, California 93106}
\author{T.~Akimoto}
\affiliation{University of Tsukuba, Tsukuba, Ibaraki 305, Japan}
\author{M.G.~Albrow}
\affiliation{Fermi National Accelerator Laboratory, Batavia, Illinois 60510}
\author{D.~Ambrose}
\affiliation{Fermi National Accelerator Laboratory, Batavia, Illinois 60510}
\author{S.~Amerio}
\affiliation{University of Padova, Istituto Nazionale di Fisica Nucleare, Sezione di Padova-Trento, I-35131 Padova, Italy}
\author{D.~Amidei}
\affiliation{University of Michigan, Ann Arbor, Michigan 48109}
\author{A.~Anastassov}
\affiliation{Rutgers University, Piscataway, New Jersey 08855}
% ...
\collaboration{CDF Collaboration}
\noaffiliation

\date{\today}

\begin{abstract}
 We report on precision measurements of the masses and widths of the narrow, orbitally excited states  
$D_1^0$ and $D_2^{*0}$ using the CDF II detector 
at the Fermilab Tevatron.   Both states (collectively called $D^{**}$) 
are reconstructed in the decay channel $D^{**}\to D^{*+} \pi^-$. The  $D_2^{*0}$ is also reconstructed in the $D^{**}\to D^{+} \pi^-$ 
channel.  Using  a dataset with an integrated luminosity of 210 $\mathrm{pb}^{-1}$, the measured 
masses and widths for the $D_1^0$ are $2421.7\pm 0.7 \pm 0.6 $ MeV/$c^2$ and
$20.0\pm 1.7\pm 1.3$ MeV/$c^2$  respectively, 
while for the $D_2^{*0}$ they are 
$2463.3\pm 0.6\pm 0.8$ MeV/$c^2$ and  $49.2\pm 2.3\pm 1.2$ MeV/$c^2$.
These values are currently the single best measurements available.

\end{abstract}


\maketitle

\begin{figure}[tbh]\centering
\hspace{-1cm}
\includegraphics[width=0.75\linewidth]{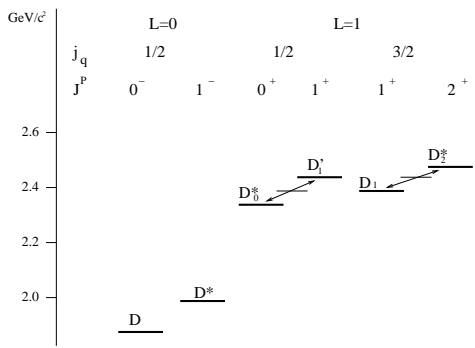}
\caption{\label{fig:massord}The mass ordering of the L=0 and L=1 D-meson 
system. The arrows  show the ``hyperfine'' splitting of the P-wave  $j_q=1/2$ and  $j_q =3/2$ states.
Properties of the broad P-wave $j_q=1/2$ states are not well established.
}
\end{figure}

The orbitally-excited charmed meson states, collectively referred to as 
$D^{**}$, are P-wave excitations of the quark-antiquark system involving 
one charm and one light quark. If we write the  total angular momentum as
$\vec{J}=\vec{j}_q+\vec{s}_Q; \vec{j}_q=\vec{s}_q+\vec{L}$, 
where $\vec{L}$ is the orbital  angular momentum and $Q (q)$ denotes the 
charm (light) quark, then in the heavy-quark limit
$m_Q \gg \Lambda_{\mathrm{QCD}}$
the spin of the charm quark $\vec{s}_Q$ decouples from the other 
degrees of freedom.  In that limit, the four P-wave states can be
separated into mass degenerate pairs:
 $j_q=1/2$ ($J^P= 0^+, 1^+$) and $j_q =3/2$ ($J^P = 1^+, 2^+$).
Heavy quark symmetry \cite{hqet} provides a systematic treatment of the
$m_Q$-dependent ``hyperfine'' splittings within each doublet, as well
as the average mass splittings between doublets, as illustrated 
in Figure~\ref{fig:massord}.  

The $D_1^0$ and $D_2^{*0}$ states, which form the  $j_q=3/2$ doublet,
are expected to decay into final states $D^*\pi$, with an overall
D-wave configuration, whereas the $j_q=1/2$ states $D_0^{*0}$ and
$D_1'{0'}$, are expected to decay in an overall S-wave
configuration.  Thus the $j_q=3/2$ states are expected to have narrow decay
widths, comparable to their mass splitting \cite{falkpesk}, while the
$j_q=1/2$ states  are expected to be much broader.
Recent theory estimations give the mass values of the two  $j_q=3/2$  states 
\cite{kala,dipier,lewis,ebert,isgur,godfrey}.

These narrow orbitally-excited charmed mesons have been observed
by several experiments \cite{argus,e687,focus,cleo,cleo90,e691,belle,d0}.  In principle, 
given the large charm cross section  at the Tevatron $p\bar{p}$
collider, very high statistics samples can be 
collected for precision measurements of the properties of these states.
However, at the trigger level it is difficult to separate low-mass fully
hadronic D-meson decays 
from the overwhelming QCD background.
The CDF II detector overcomes this obstacle with a novel two-track trigger, 
which selects long-lived charged hadrons from secondary vertices, thus
suppressing prompt charged hadrons from the QCD background.
The dataset for this analysis, based on the vertex trigger,  was taken
between March 2001 to November 2003, 
corresponding to an integrated luminosity of  $210~\mathrm{pb^{-1}}$. 
This sample is ideal for studies of charm-particle decays into two or
more hadrons.  In this analysis, both the  $D_1^0$ and $D_2^{*0}$
states have been observed in the $D^{*+} \pi^-$ channel, followed by the
decay $D^{*+}\to D^0\pi^+$ and $D^0\to K^-\pi^+$. The
$D_2^{*0}$ state has also been observed in the $D^+\pi^-$ 
channel (followed by $D^{+}\to K^- \pi^+ \pi^+$), where 
the corresponding $D_1^0$ decay is forbidden by parity and 
angular momentum conservation. Charge-conjugate decay modes are included in the analysis.

A description of CDF can be found in Ref \cite{cdf};
here only the pertinent detector components are described. 
This measurement uses  
tracks measured in the pseudorapidity
range $|\eta|<1.1$ \cite{cdfframe}, reconstructed by a silicon microstrip vertex detector 
(SVXII) \cite{svxii} and the Central Outer Tracker (COT) \cite{cot}, both in a 1.4 T solenoidal magnetic field. 
The SVXII consists of double-sided sensors
arranged in five cylindrical layers at radii between 2.5 and
10.6 cm, each providing an $r-\phi$
position measurement with a precision of $\approx 10~\mu$m.
The COT is an open-cell drift chamber with 96 layers of sense wires, 
grouped into 8 super-layers of alternating axial and $2^\circ$
stereo readout, and
providing track measurements between 40 and 137 cm in radius.

CDF collects events with a three-level trigger system.
At Level~1, two oppositely-charged tracks are reconstructed in the COT 
by the eXtremely Fast Tracker (XFT)~\cite{xft} and
are required to have a transverse momentum $p_T\ge2\,\GeVc$ each, and
$p_{T1}+p_{T2}\ge5.5\,\GeVc$. At
Level~2, the Silicon Vertex Tracker (SVT)~\cite{svt} associates SVX\,II
$r$-$\phi$ position measurements with XFT tracks, providing
a precise measurement of the track impact parameter ($d_0$),
the distance of closest approach of the track trajectory to
the beam axis in the transverse plane.
Decays of long-lived particles are identified by requiring
two tracks with $120\,\mu{\rm m}\le d_0 \le1.0\rm\,mm$,
an opening angle between the two tracks satisfying
$2^\circ\le|\Delta\phi|\le90^\circ$,
and $\Lxy
>200\rm\,\mu m$, where  $\Lxy$  is the transverse distance from the beam axis to the two track
intersection projected along the total transverse momentum of the
track pair.  A complete event reconstruction of  $D^0\to K^-\pi^+$
is performed at Level 3, where the
Level 1 and Level 2 trigger requirements are confirmed. 

\begin{figure}[tb]\centering
\includegraphics[width=0.49\linewidth]{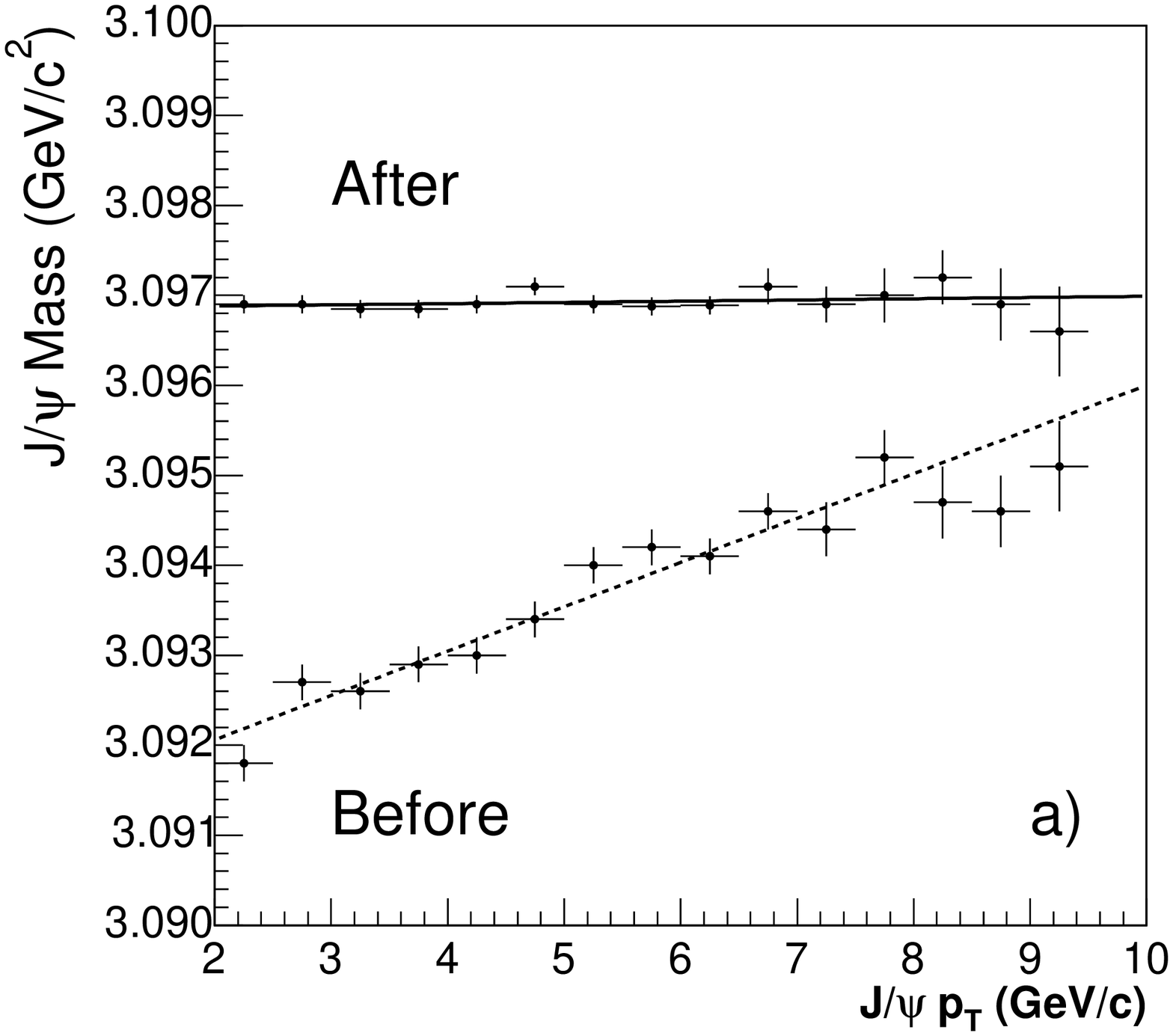}
\includegraphics[width=0.49\linewidth]{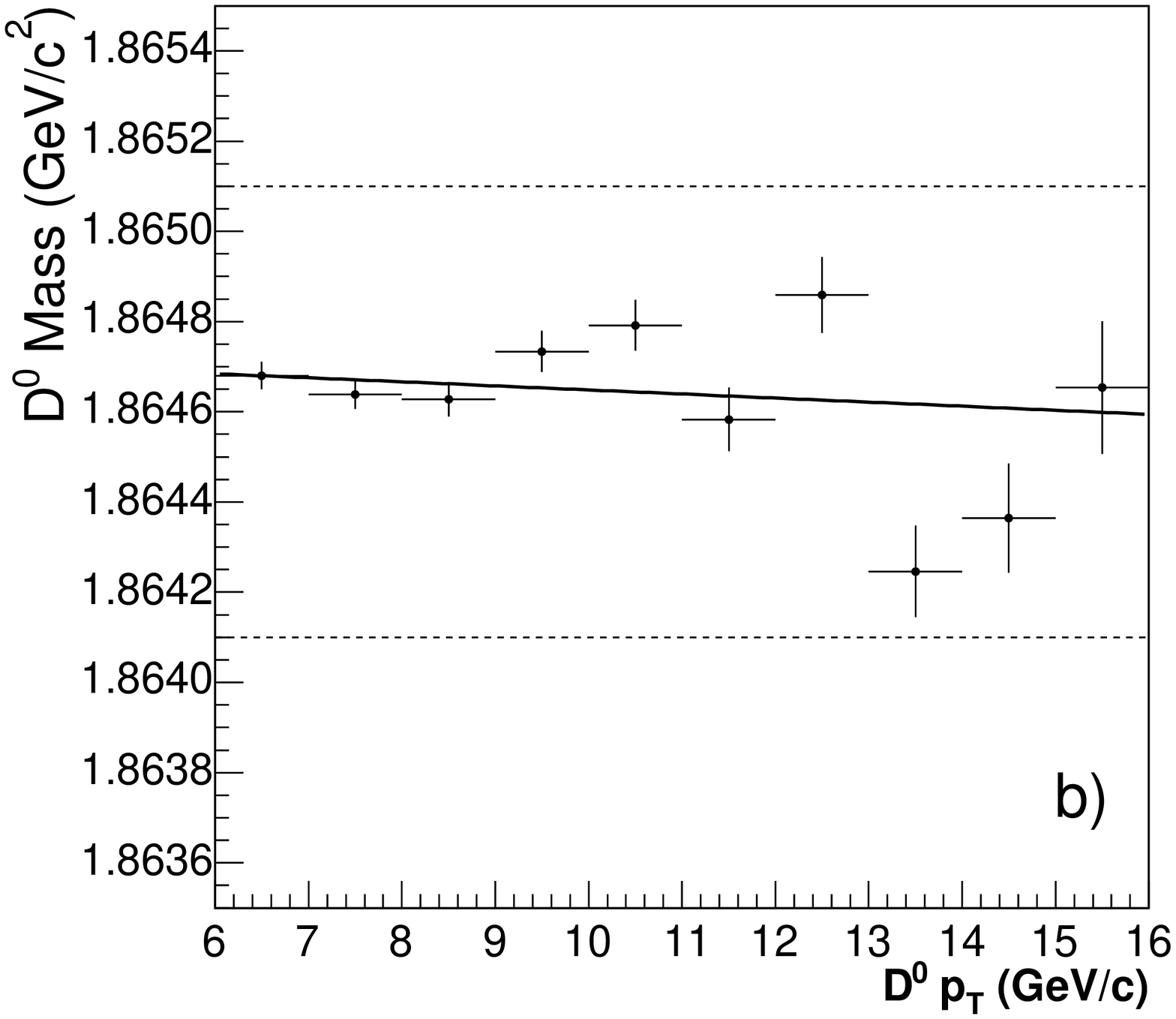}
\caption{\label{fig:massjpsi} a) The $J/\psi$ mass as a function of 
$p_T$, before (dashed) and after (solid) tracking calibration.   The slope is reduced from 0.50$\pm$
0.03 to $0.006\pm 0.010$ MeV/$c^2$ per GeV/$c$. \\b) The $D^0$ mass as a function of 
$p_T$, after calibration. The dashed lines indicate the world average
1$\sigma$ error band of $\sim 1~\mathrm{MeV}/c^2$. }
\end{figure}

The narrow width and small mass difference between the $D^{**}$ states 
require stringent tracking calibration to achieve 
adequate mass resolution.  The calibration procedure is done in two steps.
The first step is to  determine the error matrix for the COT track parameters,
accounting for multiple scattering inside the COT volume.
These uncertainties depend on the COT material description and 
hit resolution in the drift model. The error matrix is computed using
simulated $J/\psi \rightarrow \mu^+\mu^-$ decays, by analyzing pull 
distributions of the five helix parameters between generated and 
reconstructed muon tracks.  Maximum likelihood fits are performed in bins of
$p_T^2$ of the $J/\psi$ to derive the $p_T$-dependent rescaling factors.
The second step is similar to that
used in B mass measurements at CDF \cite{korn}, but with improved energy loss 
calculations.  We first use photon conversions to electron-positron
pairs to determine
the radiation length distribution of the inner detector material and then 
tune the energy loss parameters
using the reconstructed $J/\psi$ mass as a calibration. 
For this tuning we describe the detector as a series of cylindrical
layers of different materials;
in each layer we calculate the average dE/dx energy loss 
in the traversed medium and then refit the track parameters. 
The calibration uses the
$J/\psi \rightarrow  \mu^+\mu^-$ decays from data to iteratively adjust the material composition
and thickness of each layer for different regions in $z$,
until the $J/\psi$ mass dependence on the transverse momentum is negligible.  
Finally, the absolute mass scale is reached 
by adjusting the magnetic field by 0.1\% to set the
value of the $J/\psi$ mass to the world average value \cite{pdg}. 
 Figure \ref{fig:massjpsi}a shows the $J/\psi$ mass versus 
$p_T$ before and after calibration, where the slope changes from  $0.50 \pm 0.03$ MeV/$c^2$ per GeV/$c$
to  $0.006\pm 0.010$ MeV/$c^2$ per GeV/$c$ and the mean corresponds to
the world average $J/\psi$ mass. 

The tracking calibration has been cross-checked with several other particle
samples which populate different mass ranges or different decay topologies,
 such as $K_S^0$, $\Upsilon(1S)$, $B^\pm$ (to check for charge asymmetry biases) 
and $D^0$. For all cases the masses agree well with the world averages,
and the charge asymmetry was 
found to be negligible.  
The results of these tests are summarized in Table \ref{tab:xchecks}. Figure \ref{fig:massjpsi}b illustrates the results for the $D^0$,
 of particular relevance for this analysis.
\begin{table}[tb]
\centering
\begin{tabular}{|l|l|r|r|r|r|}\hline
Meson&Decay&Slope&$m - m_{\mathrm{PDG}}$ \\
&mode&(MeV/$c^2$/GeV/$c$)&(MeV/$c^2$) \\
 \hline
$K^0_s$        & $\pi^+\pi^-$   & -0.07$\pm$0.16  & -0.042$\pm$0.086 \\
$B^\pm$        & $J/\psi K^\pm$ & 0.11$\pm$0.06   & -0.27$\pm$1.037  \\
$J/\psi$       & $\mu^+\mu^-$   & -0.006$\pm$0.01 & 0$\pm$0.45        \\
$\Upsilon(1S)$ & $\mu^+\mu^-$   & -0.16$\pm$0.21  & -0.7$\pm$1.9      \\ 
$D^0$          & $K^-\pi^+$ & $(-9 \pm 6)\times 10^{-6}$& 0.1$\pm$0.55 \\
\hline
\end{tabular}
\caption{\label{tab:xchecks}Summary of mass measurements used to validate
the tracker calibration}
\end{table}

\begin{figure}[tbh]\centering
\includegraphics[width=0.85\linewidth]{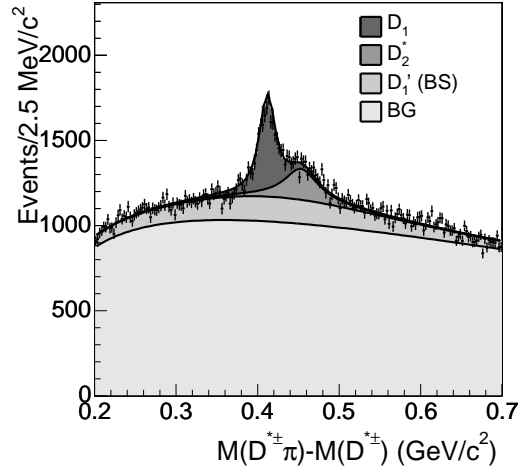}
\caption{\label{fig:d12mass} Invariant mass difference between  $D^{*}\pi$ 
system and the $D^{*}$. The points represent the data, and  lines represent the projection of the fit results for the individual components described in the text.}
\end{figure}

\begin{figure}[tbh]\centering
\includegraphics[width=0.85\linewidth]{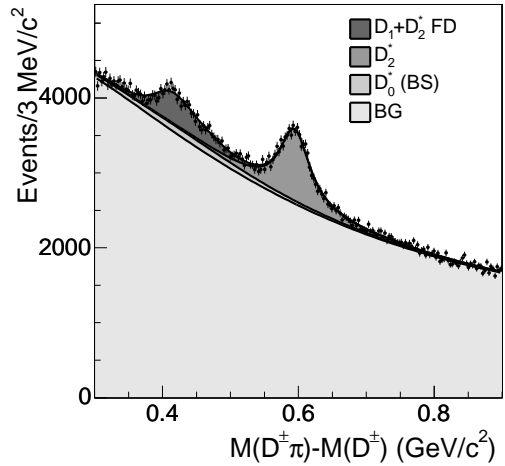}
\caption{\label{fig:d2mass} Invariant mass difference between the $D^+\pi^-$ system and the $D^+$, with the same definitions as Figure \ref{fig:d12mass}.}
\end{figure}

Events considered in this analysis are required to have at least four tracks
with  total charge zero, out of which two are consistent
with the trigger requirements ($p_T>2$ GeV/$c$ and $|d_0|>100 \mu$m). 
For the first channel ($D^{**}\to D^{*+} \pi^-$, followed by
$D^{*+}\to D^0\pi^+$ and $D^0\to K^-\pi^+$), 
these ``trigger tracks''
are required to have an invariant mass within 24 MeV/$c^2$ of the
$D^0$ mass of 1864.5 MeV/$c^2$ \cite{pdg}. Particles are assigned in turn the
$K$ and $\pi$ masses, and all $K \pi$ combinations falling 
within the mass window are retained, to avoid any bias in the $D^0$ mass.
This $D^0$ candidate is associated with another track with $p_T > 400$ MeV/$c$
to form a $D^{*+}$
candidate, with the requirement that the mass difference between the $D^{*+}$ and the
$D^0$ be smaller than 147 MeV/$c^2$. 
In addition, to reduce both background and misassignment of the $K$ and
$\pi$ masses, the third track is
required to have the same charge as the candidate pion track, as
expected from leading order $D^{*+}$ decays.
Finally, the $D^{*+}$ is associated
with a negative track with $p_T > 400$ MeV/$c$, to form a four-track 
$D^{**}$ candidate.
For the second channel ($D_2^{*0} \to D^+\pi^-$, followed by $D^{+}\to K^- \pi^+ \pi^+$), no mass window is required for the 
trigger tracks. The $D^+$
candidate is constructed from three tracks with a mass and charge 
assignment compatible with being a $K^- \pi^+\pi^+$ system; two of
these tracks must satisfy the trigger requirements, while the third must have
$p_T > 800$ MeV/$c$.  We first ensure that all three tracks originate
from a common vertex which is well separated from the primary vertex, by 
requiring the $\chi^2$ from a 3-dimensional fit to the three
tracks to be smaller than 12, and have an associated $L_{xy} > 1$ mm. 
Then, $D^+$ candidates are defined as three-track systems with invariant mass between 1.85 and 1.89 GeV/$c^2$.
Finally, to obtain a neutral $D^{**}$ candidate, the three tracks are combined
with a fourth, of opposite charge with respect to the sum of the first three. 
The two $D^{**}$ resonances are analyzed in terms of the invariant mass
difference between the four-track and the three-track system, which
is crucial to separate the resonances from the background. 
Figure
\ref{fig:d12mass} and \ref{fig:d2mass} show the results for the
$D^{**}$ resonances in the two decay channels. 

In Figure \ref{fig:d12mass}, signals from the $D_1^0$ at 0.40 GeV/$c^2$ 
and the $D_2^{*0}$ at 0.46 GeV/$c^2$ are
clearly visible above the combinatorial background.
Due to the unknown shape of the combinatorial background it is not possible
with these data to derive with sufficient confidence the properties, or even the
existence of the broad state (BS)  $D_1^{0'}$ (the $1^+$ component of
the $j_q=1/2$ doublet) from our data.
Since a state with width around 200 MeV/$c^2$ is suggested by
a Belle measurement \cite{belle} its potential effect is included as
a systematic error.

The distribution of the mass difference in Figure \ref{fig:d2mass} presents 
two features: a broad peak around 0.4 GeV/$c^2$, and
a narrow pronounced peak around 0.6 GeV/$c^2$. The broad peak is due to
the feed-down decays of the type $D^{**}\to
D^{*+}\pi^-$ accessible for both $D_1^0$ and $D_2^{*0}$, followed by $D^{*+}\to D^+ \pi^0$, where the
$\pi^0$ is not observed.  Similar to the  $D_1^{0'}$ in Figure
\ref{fig:d12mass}, part of this peak could 
be due to a broad neutral $D^{*0}_0$ state with a mass around
2410 MeV/$c^2$ and width of about 250 MeV/$c^2$, as reported by FOCUS \cite{focus}.
The peak around 0.6 GeV/$c^2$ is the main $D_2^{*0}$ signal.

A simultaneous binned likelihood fit to both histograms in Figures \ref{fig:d12mass} and
\ref{fig:d2mass} is used to extract the mass and width for the $D_1^0$ and $D_2^{*0}$.
The likelihood function consists of:
\begin{itemize}
\item[i) ] a signal term for each narrow state in the $D^{*+}$ channel, and only
a $D_2^{*0}$ term in the $D^+$ channel. These terms are a convolution of a
non-relativistic Breit-Wigner (BW) distribution with a resolution histogram taken from
Monte Carlo simulation, and depends on the amplitudes $A_{1,2}$, the mass
differences with respect to the three-track states
$M_{1,2}$ and widths $\Gamma_{1,2}$. The $D_2^{*0}$ widths are common for the 
two channels. The likelihood component is:
\[f_{D_{1,2}}(\Delta m) = A_{1,2}\times BW(\Delta m, M_{1,2}, \Gamma_{1,2})\otimes Resolution. \]
The $D_2^{*0}$ masses in the two channels are connected
by using the world average value for the mass difference between $D^{*+}$ and $D^*$.
\item[ii) ] a background (BG) term for each histogram with free 
parameters $\alpha$, $\beta$, $\gamma$ and $\delta$ (set to zero for the first channel), of the form:
\[f_{BG}(\Delta m) = \alpha (\Delta m-m_\pi)^\beta 
e^{-\gamma(\Delta m-m_\pi)}+\delta.\]
\item[iii) ] a broad state (BS) term for each histogram, modeled as 
a Breit-Wigner function similar to that of the narrow resonances, but convoluted with a 
Gaussian whose width is taken from simulation:
\[f_{BS}(\Delta m) = A_{BS} \times BW(\Delta m, M_{BS}, 
\Gamma_{BS})\otimes Resolution.\]
\item[iv) ] a feed-down (FD) term for  decays of $D_1^0$ and $D_2^{*0}$ to $D \pi^0 
\pi$, where the $\pi^0$ is lost and a smaller mass is reconstructed. The 
only additional free parameter is a common scale factor relative to the 
amplitudes of the $D^{*}\pi$ channel, since the  $D_1^0$ and 
$D_2^{*0}$ masses and widths are the same. The shift and resolution
are taken from Monte Carlo.

\end{itemize}
Overall we have 7 parameters for the narrow resonances (amplitude, mass and
width for each each resonance, plus the $D_2^{*0}$ yield in the second channel), 7 for background
modeling (one parameter is set to zero in the first channel), two amplitudes for
broad states and one feed-down normalization, for a total of 17 parameters. If the mass
and width of one or two broad states are left floating, we can have up to 19 or
21 parameters.
In order to properly describe the tails of the resolution, 
avoiding detector biases which occur when a Gaussian
approximation is made, we use the explicit resolution histogram
for signal and feed-down. The detector resolution is about 4 
MeV/$c^2$, much smaller than the intrinsic width of the 
resonances, but the tails are significantly larger.
On the other hand, Gaussian resolution is sufficient for the broad states, 
since the tails fall outside the fitting window.
The accuracy of the simulation of the detector resolution has been tested in 
a control sample of $D^0$ and $D^{*+}$, whose natural width is negligible 
with respect to detector resolution.  After taking the detector resolution 
into account using the above
prescription, we measure a width smaller than 0.2 MeV/$c^2$ for all momentum 
ranges, and use this value as a conservative systematic 
error on the width due to tracking precision. If instead, the Gaussian 
approximation was made for the width, we then observed width and mass shifts of the order
of a few MeV$/c^2$, much larger than the precision envisaged.

The fitting procedure has been tested on a Monte Carlo sample of fully
simulated $D_1^0\to D^{*+}\pi^-$, $D_2^{*0}\to D^{*+}\pi^-$ and $D_2^{*0}\to D^+\pi^-$ 
decays with three times the statistics of the observed data. The feed-downs
are also included. However, they have a small impact on the final result.
Both background and broad-state events 
have been generated from the parametrized data distribution.  With
this sample,
two important cross-checks on the fitting algorithm have been performed.
First, the full simulation sample is fitted and the results agree with the 
simulation input parameters.  In addition, a large number of ``toy
experiments'' with statistics comparable to the data are generated
and fit to look for potential biases.  The resultant pull distribution
for each fit parameter is consistent with a Gaussian distribution of zero 
mean and unit width.

The likelihood fit applied to the data
yields roughly 7500 $D_1^0$ and 5000 $D_2^{*0}$ candidates for the channel 
of Figure \ref{fig:d12mass} and
20000 $D_2^{*0}$ candidates for that of Figure \ref{fig:d2mass}.
Given the large background of unknown shape, there is no way to derive the
presence of a broad state from our data only. Leaving all its parameters
free, the resulting broad state yield is
small and the $\chi^2$ quality of fit probability is 51\%. Three additional hypotheses on the broad
state are tested: a mass and width as measured by FOCUS \cite{focus}, as measured by  
Belle \cite{belle}, or omitting the BS term from the fit.  
The resulting variations in the measured quantities are shown in Table \ref{tab:bs}.
\begin{table}[tbh]
\centering
\hspace{0pt}\begin{tabular}{|l|l|l|l|l|l|}\hline
Model & $\Delta (M_{D_1^0})$ & $\Delta(\Gamma_{D_1^0})$ & $\Delta (M_{D_2^{*0}})$&$\Delta(\Gamma_{D_2^{*0}})$&$P(\chi^2)$\\
\hline
Free&---&---&---&---&51\\
FOCUS&0.1&0.2&0.1&0.5&50\\
Belle&0.1&0.3&0.1&0.5&50\\
Absent&0.1&0.4&0.1&0.6&52\\
\hline
\end{tabular}
\caption{\label{tab:bs} Shift of the measured masses and widths of
 $D_1^0$ and $D_2^{*0}$ and  $\chi^2$ probability P for different
hypotheses of the broad state; allowing parameters to float in the
 fit,  using the values published by FOCUS
 \cite{focus} or Belle \cite{belle},
or forcing it to zero in the $D \pi$ channel.  Mass shifts are in
 $\MeVcc$ and probability in \%.} 
\end{table}
As the differences are small and there is no discrimination on the
basis of the $\chi^2$ probability, the central result includes the
broad state with free parameters, and other options
provide estimates of systematic uncertainty.
\begin{table}[h]
\begin{center}
\begin{tabular}{|l|c|c|c|c|}
\hline%
Source & $\Delta (M_{D_1^0})$ & $\Delta(\Gamma_{D_1^0})$ & $\Delta (M_{D_2^{*0}})$&$\Delta(\Gamma_{D_2^{*0}})$\\
\hline
MC statistics& 0.3&1.2&0.4&1.2\\ \hline
Broad State& 0.1&0.4&0.1&0.5\\
Track Error scale & 0.1&---&0.1&---\\
Fit model & $<0.1$ & $<0.1$ & $<0.1$ & $<0.1$\\
Mass Calibration & 0.1&0.2&0.1&0.2\\ \hline
Total (Relative) & 0.4 & 1.3 & 0.5&1.3\\ \hline
Reference mass & 0.5 & --- &0.7 & ---\\ \hline 
Total (Absolute) & 0.6&1.3&0.8&1.3\\ \hline 
\end{tabular}
\end{center}
\caption{\label{tab:syst}Summary of  systematic error for the mass and width of the two
resonances, all in units of $\MeVcc$. Since the measurement is a mass difference,
the absolute mass error includes a systematic error
due to the uncertainty of the mass of the reference
particle. }
\end{table}

From the likelihood fit we derive results, with statistical uncertainties only,
 on the $D_1^0$ and $D_2^{*0}$ widths and mass differences with respect to the $D^{*+}$ or $D^+$ into which the $D^{**}$ decays.
To derive absolute masses from the measured mass differences
we add the world average masses \cite{pdg} of the $D^{*+}$
or $D^+$, and the corresponding uncertainty on this value (both known
with 0.5 MeV/$c^2$ precision) to the systematic errors.
The remaining residual systematic errors are connected with Monte Carlo statistics
and the tracking calibration, and are listed in Table \ref{tab:syst}.

\begin{figure}[tbh]\centering
\includegraphics[width=0.85\linewidth]{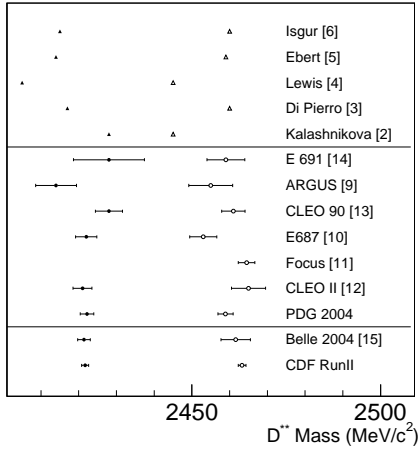}
\caption{\label{fig:theory} $D_1^0$ (solid) and $D_2^{*0}$ (open) mass
  comparison with theoretical expectations (triangles) and previous
  measurements and current world average (circles). Errors for theory predictions
are not readily available.}
\end{figure}

In summary, the large sample of orbitally-excited charmed mesons collected by 
the CDF collaboration
using the vertex trigger allows the measurement of properties of
the L=1 orbitally-excited narrow states $D_1^0$ and $D_2^{*0}$ with unprecedented
precision.
Using both $D^{*+}\pi^-$ and $D^+\pi^-$ final states, the measured widths of
these states are:
\begin{eqnarray}
\nonumber \Gamma(D_1^0) &=& 20.0\pm 1.7 \pm 1.3 ~\mathrm{MeV/}c^2, \\
\nonumber \Gamma(D_2^{*0}) &=& 49.2\pm 2.3\pm 1.3 ~\mathrm{MeV/}c^2.
\end{eqnarray}
where the first error is statistical, and the second systematic, above
and for all results which follow. In order to improve the 
resolution, the mass measurement is performed in term of 
differences with respect to the daughter 
charmed particle. The measured mass differences are:
\begin{eqnarray}
\nonumber M(D_1^0)-M(D^{*+}) &=& 411.7\pm 0.7\pm 0.4 ~\mathrm{MeV/}c^2, \\
\nonumber M(D_2^{*0})-M(D^{+}) &=& 593.9\pm 0.6\pm 0.5 ~\mathrm{MeV/}c^2.
\end{eqnarray}
The $D_2^{*0}$ mass is given with respect to the $D^+$ since almost all the
information comes from this decay channel.
Adding the world average values for the $D^{*+}$ and $D^+$ masses and including their 
uncertainties in quadrature to the systematic error, the absolute values
of the masses are:
\begin{eqnarray}
\nonumber M(D_1^0) &=& 2421.7\pm 0.7\pm 0.6~\mathrm{MeV/}c^2,\\
\nonumber M(D_2^{*0}) &=& 2463.3\pm 0.6\pm 0.8~\mathrm{MeV/}c^2.
\end{eqnarray}
A comparison of these values with the world averages and some recent theoretical models can be 
found in Figure \ref{fig:theory}; the results are consistent with the models, and with
better estimations of uncertainties from the models they should allow one to
constrain which theoretical picture is consistent with experiment. 
Comparison with predictions and measurements for the widths is less insightful due
to the larger uncertainties; however, the width measurements presented
here are consistent with other recent measurements \cite{belle,cleo} .

This is the best single world measurement of the
masses of the orbitally-excited charm states. The 
total error is still limited by the statistics, and there is room for
improvement with the data presently being recorded by the CDF II experiment.
A precise determination of the theory uncertainties would however be needed
to assess the discriminating power of this and future measurements.

We thank the Fermilab staff and the technical staffs of the participating 
institutions for their vital contributions. This work was supported by the 
U.S. Department of Energy and National Science Foundation; the Italian 
Istituto Nazionale di Fisica Nucleare; the Ministry of Education, Culture, 
Sports, Science and Technology of Japan; the Natural Sciences and Engineering 
Research Council of Canada; the National Science Council of the Republic of 
China; the Swiss National Science Foundation; the A.P. Sloan Foundation; the 
Bundesministerium f\"ur Bildung und Forschung, Germany; the Korean Science 
and Engineering Foundation and the Korean Research Foundation; the Particle 
Physics and Astronomy Research Council and the Royal Society, UK; the Russian 
Foundation for Basic Research; the Comisi\'on Interministerial de Ciencia y 
Tecnolog\'{\i}a, Spain; in part by the European Community's Human Potential 
Programme under contract HPRN-CT-2002-00292; and the Academy of Finland. 

%


\begin{thebibliography}{00}
\bibitem{cdfframe} CDF uses a cylindrical coordinate system ($r,\phi,z$) with the
 origin at the center of the detector and the 
 positive $z$ direction aligned with the proton direction.
 Psuedorapidity $\eta$ is defined as $\eta \equiv -\ln(\tan(\theta/2))$, where $\theta$ is the polar angle.
\end{thebibliography}
\end{document}